\documentclass[%
 aip,
 amsmath,amssymb,
 reprint,%
]{revtex4-1}

\usepackage{graphicx}% Include figure files
\usepackage{dcolumn}% Align table columns on decimal point
\usepackage{bm}% bold math
\usepackage[utf8]{inputenc}
\usepackage[T1]{fontenc}
\usepackage{mathptmx}
\usepackage{etoolbox}
\usepackage[colorlinks=true,linkcolor=blue,citecolor=blue]{hyperref}
\makeatletter
\def\@email#1#2{%
 \endgroup
 \patchcmd{\titleblock@produce}
  {\frontmatter@RRAPformat}
  {\frontmatter@RRAPformat{\produce@RRAP{*#1\href{mailto:#2}{#2}}}\frontmatter@RRAPformat}
  {}{}
}%
\makeatother
\begin{document}

\preprint{AIP/123-QED}

\title[]{Realization of the unidirectional amplification in a cavity magnonic system}
% Force line breaks with \\
\author{Zi-Yuan Wang}
\affiliation{Interdisciplinary Center of Quantum Information, State Key Laboratory of Extreme Photonics and Instrumentation, and Zhejiang Province Key Laboratory of Quantum Technology and Device, School of Physics, Zhejiang University, Hangzhou 310027, China}%Lines break automatically or can be forced with \\
\author{Jie Qian*}%
 \email{qjie@zju.edu.cn}
\affiliation{Interdisciplinary Center of Quantum Information, State Key Laboratory of Extreme Photonics and Instrumentation, and Zhejiang Province Key Laboratory of Quantum Technology and Device, School of Physics, Zhejiang University, Hangzhou 310027, China}%

\author{Yi-Pu Wang*}
\email{yipuwang@zju.edu.cn}
\affiliation{Interdisciplinary Center of Quantum Information, State Key Laboratory of Extreme Photonics and Instrumentation, and Zhejiang Province Key Laboratory of Quantum Technology and Device, School of Physics, Zhejiang University, Hangzhou 310027, China}%

\author{Jie Li}
\affiliation{Interdisciplinary Center of Quantum Information, State Key Laboratory of Extreme Photonics and Instrumentation, and Zhejiang Province Key Laboratory of Quantum Technology and Device, School of Physics, Zhejiang University, Hangzhou 310027, China}%

\author{J. Q. You}%
\affiliation{Interdisciplinary Center of Quantum Information, State Key Laboratory of Extreme Photonics and Instrumentation, and Zhejiang Province Key Laboratory of Quantum Technology and Device, School of Physics, Zhejiang University, Hangzhou 310027, China}%

\date{\today}% It is always \today, today,

\begin{abstract}
We experimentally demonstrate the nonreciprocal microwave amplification using a cavity magnonic system, consisting of a passive cavity (i.e., the split-ring resonator), an active feedback circuit integrated with an amplifier, and a ferromagnetic spin ensemble (i.e., a yttrium-iron-garnet sphere). Combining the amplification provided by the active circuit and the nonreciprocity supported by the cavity magnonics, we implement a nonreciprocal amplifier with the functions of both unidirectional amplification and reverse isolation. The microwave signal is amplified by 11.5~dB in the forward propagating direction and attenuated in the reverse direction by -34.7~dB, giving an isolation ratio of 46.2~dB. Such a unidirectional amplifier can be readily employed in quantum technologies, where the device can simultaneously amplify the weak signal output by the quantum system and isolate the sensitive quantum system from the backscattered external noise. Also, it is promising to explore more functions and applications using a cavity magnonic system with real gain.
\end{abstract}

\maketitle

Nonreciprocal devices, which exhibit the characteristic of unidirectionality, can play a crucial role in the information technology. To realize the nonreciprocity, various methods have been proposed, such as the magneto-optical
Faraday rotation~\cite{Wolfe-95,Wang-05,Belotelov-07,Bi-11,Chin-13}, nonlinearity~\cite{Christodoulides-88,Shi-15,Fan-11,Wang-13,Huang-18}, spatial-temporal modulation~\cite{Hwang-97,WangDW-13,Doerr-11,Lira-12}, and reservoir engineering~\cite{Metelmann-15,Fang-17}. The main function of the nonreciprocal devices is to both transmit information in the desired direction and isolate the backscattered noise. The typical feature is that the signal is transmitted in one direction with some insertion loss, and the reverse direction propagation has a large attenuation. On the other hand, in the process of signal transmission, the signal will inevitably attenuate continuously when increasing the propagation distance, so the gain medium and amplification device~\cite{Xiao-10,Massel-11,Fermann-00,Leon-10,Liu-07,Khurgin-12,Choksi-22,Stehlik-16,Frolov-99,Yao-17,Yao-23} are also a critical component in the information technology.

Under normal circumstances, the nonreciprocal and gain devices are independent of each other in the information network, each performing its own function separately. With the increasingly high requirements for integrating signal processing devices, whether both nonreciprocity and signal amplification can be realized in the same device has become an urgent demand. Different systems were proposed to realize the unidirectional amplification, including the atomic system~\cite{Lin-19}, whispering-gallery microcavities~\cite{Peng-14}, Josephson circuit~\cite{Abdo-13,Abdo-14}, microwave optomechanical device~\cite{Malz-18,Mika-19}, magnonic system~\cite{Zhao-22A}, and others~\cite{Koutserimpas-18,Kamal-17,Jiang-18,Galiffi-19,Song-19,Taravati-21}.

Recently, cavity magnonics  has emerged as a new research frontier~\cite{Rameshti-14}. Based on this platform, various applications in, e.g., memory~\cite{Zhang-15,Shen-21}, cavity optomagnonics for quantum transduction~\cite{Osada-16,Tabuchi-15,Hisatomi-16,James-15,James-16,Tang-16,Kusminskiy-16,Kusminskiy-18,Bauer-19,Weijiang-21,Lauk-20}, and magnon sensing at the quantum level~\cite{Quirion-20,Wolski-20,Xu-22,xu2023deterministic} have been reported. In the cavity magnonic system, a magnon mode (i.e., the Kittel mode) in a ferrite, e.g., the yttrium-iron-garnet (YIG) sphere, strongly couples to the microwave photons in the cavity. As a mature commercial material, the YIG owns significant advantages such as the high spin density~\cite{Hans-13,Zhang-14,Tabuchi-14}, flexible adjustment of resonant frequency~\cite{Chumak-15,WangY-20}, and low damping rate. Very recently, novel properties of the nonreciprocity and unidirectional invisibility have also been observed in the cavity magnonic system~\cite{WangY-19}, which arise from the interference between the coherent and dissipative magnon-photon interactions. By combining the easy-to-tune ferrite and the specially designed cavities, isolators with both nearly infinite isolation ratios and adjustable working frequencies become implementable~\cite{WangY-19,Qian-20,Xufeng-20,Zhao-20,Chai-21,Kim-22}.

In this work, we construct a nonreciprocal amplifier based on the cavity magnonic system, consisting of a passive cavity made by the split-ring resonator (SRR), an active circuit with an embedded amplifier, and a YIG sphere. These components are all connected to a strip-line waveguide that supports the traveling photon mode. The active circuit is employed to compensate the dissipation of the passive cavity until the signal amplification occurs. The coupling strength between the magnon and cavity modes can be modified by adjusting the position of the YIG sphere with respect to the passive cavity. Through the comprehensive and synergistic regulation of the active circuit and magnon-photon coupling strength, we experimentally realize the microwave nonreciprocal amplification with a high isolation ratio. In the optimal conditions, the device exhibits a rightward propagating amplificition of 11.5~dB and a reverse propagating attenuation of -34.7~dB. Such devices have potential applications in quantum networks and repeaters~\cite{Rempe-15,Hanson-18,Zhong-21,Nguyen-19,Li-21}. They may also be promising in protecting the sensitive quantum system from noises associated with the read-out electronics and amplifying the weak signal leaking out of the quantum nodes.

%\section{\label{sec:level1}SYSTEM AND MODEL}

\begin{figure*}[t]
	\includegraphics[width=0.78\textwidth]{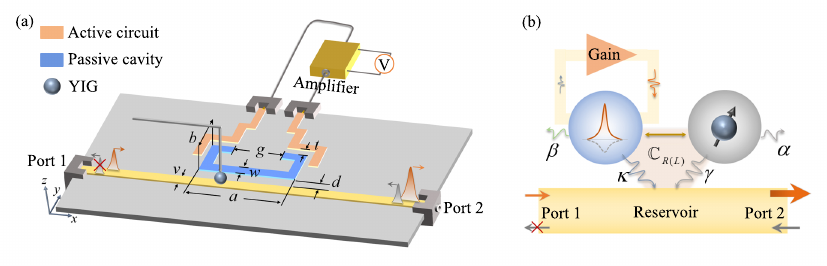}
	\caption{(a) Schematic diagram of the experimental setup, consisting of a YIG sphere glued at the end of a displacement cantilever, a passive cavity made by the split-ring resonator (SRR), and an active circuit with an embedded amplifier. The DC voltage (V) applied on the amplifier is used to sustain the gain in our device. (b) Schematic diagram of the cooperative dissipation as well as the complex coupling between the cavity mode and the magnon mode.\label{fig1}}
\end{figure*}

The nonreciprocal amplification device is depicted in Fig.~\ref{fig1}(a), where the microwave circuit consists of a strip-line waveguide, a passive cavity made by the split-ring resonator (SRR), and an active circuit with an embedded amplifier. For the passive cavity, a SRR wih parameters $g=18$ mm, $b=5.2$ mm, $a=25$ mm, and $w=1.5$ mm is side-coupled to the strip-line waveguide with a distance of $d=0.2$ mm, and the resonant frequency of the SRR is designed to be $\omega_{\rm{c}}/2\pi=3.03$ GHz to match the optimal working frequency of the active circuit~\cite{Zarifi-15}. The active circuit is capacitively coupled to the SRR via a gap of $t=0.8$ mm. By introducing a gain through the amplifier, the active circuit can compensate the loss of the passive cavity. In the experiment, we apply a DC voltage on the amplifier to continuously adjust the gain provided by the active circuit. The strip-line waveguide has a width of $v=2.53$ mm for 50 $\Omega$ impedance matching. Both the cavity and waveguide are fabricated on a F4B substrate. A 1~mm-diameter YIG sphere is glued at the end of a displacement cantilever, and the relative position between the YIG sphere and the planar microwave circuit can be finely adjusted by a three-dimensional (3D) motor-controlled robot arm. An external magnetic field is applied perpendicularly to the planar device, which is used to both saturate the magnetization of the YIG sphere and tune the frequency of the magnon mode. To characterize this device, a vector network analyzer (VNA) with a signal power of -20 dBm is used to measure the transmission spectrum $|S_{12(21)}|$, where the subscript 21 (12) indicates that the microwave field is loaded to the port 1 (2) and propagates to the port 2 (1) {(see Supplementary Material Sec.~I)}.

Our device is schematically shown in Fig.~\ref{fig1}(b), where the cavity mode and the magnon mode dissipate {\it cooperatively} to the traveling wave-type dissipative reservoir (waveguide). Here, $\alpha$ and $\beta$ are the intrinsic damping rates of the magnon mode and the passive cavity mode, respectively, and $\beta$ can be reduced by the active circuit, or even turned to be negative. The external damping rates of the magnon mode and the cavity mode, i.e., $\gamma$ and $\kappa$, reflect their interactions with the waveguide traveling photon modes. The cooperative dissipation gives rise to the dissipative coupling between the cavity mode and the magnon mode~\cite{Metelmann-15,Fang-17,WangY-19}. Meanwhile, the cavity mode and the magnon mode can directly interact with each other via the spatial mode overlapping, which is attributed to the coherent magnon-photon coupling.

When the coupling mechanism of our system is dominated by both coherent and dissipative couplings, the interference between them has a significant modulation on the transmission spectrum of the system. Due to the asymmetric location of the YIG sphere with respect to the central line of the device, the phase difference between the coherent and dissipative couplings will be different when the microwave is loaded to the ports 1 and 2. Under the circumstances, the overall coupling between the cavity mode and the magnon mode can become {\it direction-dependent}. In other words, the nonreciprocity stems from the inconsistent interference effect between the coherent and dissipative couplings when the microwave transmission direction is reversed~\cite{WangY-19}.

We define $\mathbb{C}_{R(L)}=M+iN$ to characterize the direction-dependent complex coupling, where $M$ denotes the coherent coupling, $iN$ corresponds to the dissipative coupling, with $i$ indicating the non-Hermiticity of the dissipative coupling, and the subscript $R(L)$ represents the case when the signal is loaded to the port 1 (2). When the traveling wave propagates in the opposite direction, $\mathbb{C}_{R}\neq \mathbb{C}_{L}$, so the nonreciprocity emerges. Under the rotating-wave approximation, the Hamiltonian of the non-Hermitian system can be written as
\begin{equation}\label{eq0}
	H/\hbar=\tilde{\omega}_{\rm{c}}a^{\dag}a+\tilde{\omega}_{\rm{m}}b^{\dag}b+\mathbb{C}_{R(L)}(a^{\dag}b+b^{\dag}a),
\end{equation}
where $\tilde{\omega}_{\rm{c}}=\omega_{\rm{c}}-i(\beta+\kappa)$ and $\tilde{\omega}_{\rm{m}}=\omega_{\rm{m}}-i(\alpha+\gamma)$ are the complex frequencies of the cavity mode and the magnon mode, respectively, while $a^{\dag}$ ($a$) and $b^{\dag}$ ($b$) are the creation (annihilation) operators of the cavity mode and the magnon mode, respectively.

\begin{figure}[t]
	\centering
	\includegraphics[width=0.49\textwidth]{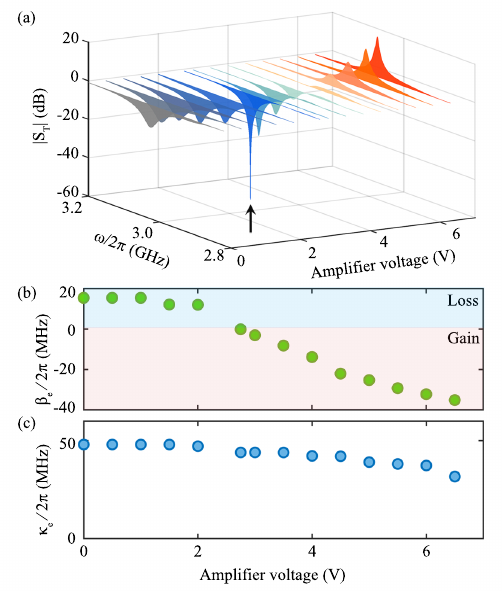}
	\caption{(a) The measured transmission spectra $|S_{\rm{T}}|$ versus the amplifier bias voltage $V$. The arrow labels the critical point, where the intrinsic dissipation is exactly compensated by the active circuit. (b) and (c) The effective intrinsic damping rate $\beta_{\rm{e}}$ and the effective external damping rate $\kappa_{\rm{e}}$ of the cavity mode versus the amplifier bias voltage $V$. In the blue region of (b), the cavity is lossy, while in the red region, the intrinsic loss of the cavity mode is compensated.}\label{fig2}
\end{figure}

\begin{figure}[t]
\includegraphics[width=0.49\textwidth]{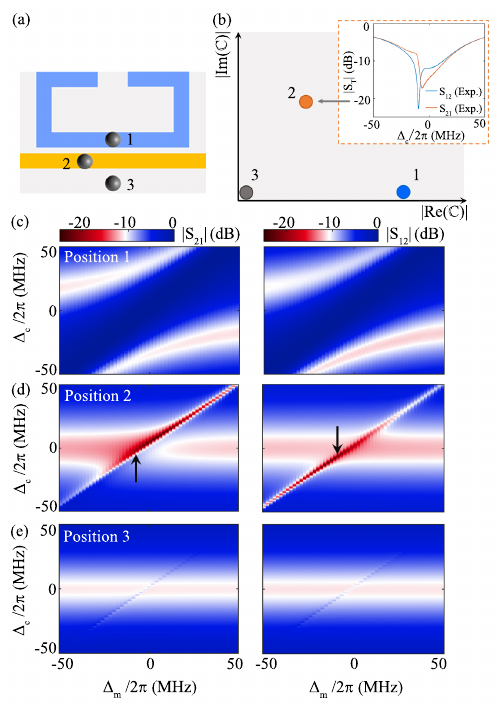}
\caption{(a) Schematic diagram of three different positions (i.e., positions 1, 2, and 3), at which the YIG sphere is placed, respectively, in the experiment. (b) Complex coupling strength obtained when the YIG sphere is placed at these three positions, respectively. (c)-(e) Measured rightward ($|S_{21}|$) and leftward ($|S_{12}|$) transmission mappings versus the field detuning $\Delta_{\rm{m}}=\omega-\omega_{\rm{m}}$ and the probe frequency detuning $\Delta_{\rm{c}}=\omega-\omega_{\rm{c}}$, when the YIG sphere is placed at positions 1, 2, and 3, respectively. The measured transmission spectra $|S_{21}|$ and $|S_{12}|$ at the arrow-marked fields in (d) are plotted as the inset panel in (b), which shows a clear feature of nonreciprocal transmission.}\label{fig3}
\end{figure}

Below we first characterize the loss compensation of the active circuit to the passive cavity, and find out the relationship between the bias voltage of the amplifier and the compensated dissipation of the passive cavity. Figure~\ref{fig2}(a) shows the response of the cavity mode to the bias voltage applied on the amplifier, where the measured transmission spectra $|S_{\rm{T}}|$ at a series of bias voltages are plotted. When the voltage is zero, $|S_{\rm{T}}|<0$, and the spectrum shows a dip at the resonant frequency of the cavity mode. With the increase of the amplifier voltage, the amplitude of $|S_{\rm{T}}|$ decreases until an ultra-sharp dip ($\sim$55 dB) appears. Then, it gradually increases and the spectrum turns into a resonance peak with $|S_{\rm{T}}|>0$. This process is accompanied with the intrinsic damping rate of the cavity mode compensated to be zero and then negative. The transmission coefficient of the side-coupled cavity mode can be described by the following equation:
\begin{equation}\label{eq1}
S_{\rm{T}}=1+\frac{\kappa_{\rm{e}}}{i(\omega-\omega_{\rm{c}})-(\kappa_{\rm{e}}+\beta_{\rm{e}})},
\end{equation}
where $\kappa_{\rm{e}}=\kappa\cdot\eta$ and $\beta_{\rm{e}}=\beta-g$ are the effective external and intrinsic damping rates of the cavity mode, respectively. The parameter $\eta$ indicates that the external damping rate of the cavity mode is affected by the active circuit when the bias voltage is applied. The parameter $g$ is introduced as the gain coefficient, which is determined by the amplifier bias voltage and is equal to the linewidth compensation of the cavity mode.

\begin{figure*}[t!]
	\centering
	\includegraphics[width=0.88\textwidth]{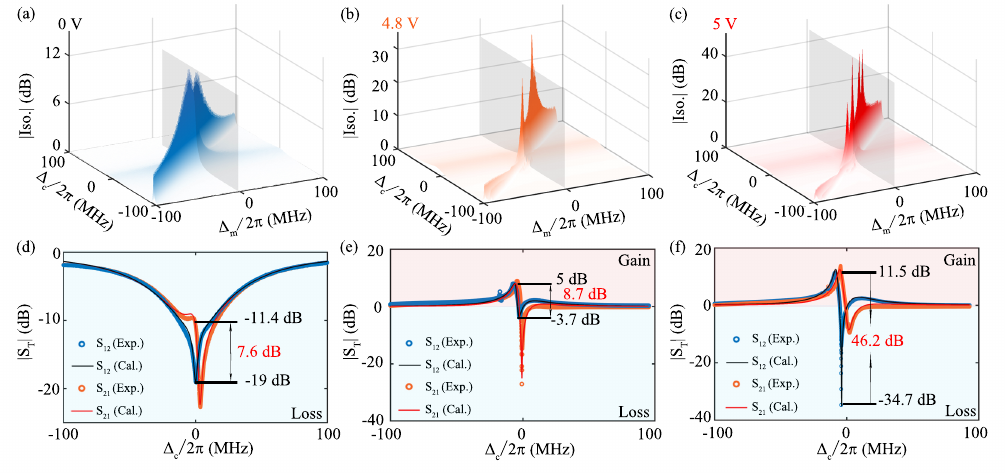}
	\caption{(a)-(c) Measured isolation ratio versus $\Delta_{\rm{m}}$ and $\Delta_{\rm{c}}$, where the amplifier bias voltage is set to be 0, 4.8 and 5~V, respectively. (d)-(f) Experimental (circles) and theoretical (solid curve) results of the $|S_{21}|$ and $|S_{12}|$ spectra at the field detuning marked by the gray plane in (a), (b) and (c), respectively. When $|S_{\rm{T}}|>0$, the propagating microwave signal is amplified. On the contrary, when $|S_{\rm{T}}|<0$, the microwave signal is attenuated. (d) A 7.6~dB isolation ratio is realized without amplification. (e) A 8.7~dB isolation ratio is realized with rightward 5~dB amplification and leftward 3.7~dB attenuation. (f) A 46.2~dB isolation ratio is realized with rightward 11.5~dB amplification and leftward 34.7~dB attenuation.}
\label{fig4}
\end{figure*}

The bare cavity mode (corresponding to zero bias voltage) has an external damping rate of $\kappa/2\pi=48$ MHz, and an intrinsic damping rate of $\beta/2\pi=15.5$ MHz. By continuously tuning the bias voltage, the fitted $\beta_{\rm{e}}$ and $\kappa_{\rm{e}}$ versus the bias voltage are plotted as the green and blue circles in Fig.~\ref{fig2}(b) and Fig.~\ref{fig2}(c), respectively. When the amplifier voltage is small, the intrinsic dissipation of the cavity mode is not fully compensated ($g<\beta$, $\beta_{\rm{e}}>0$). While the amplifier voltage reaches 2.75 V, the intrinsic damping rate of the cavity mode is almost exactly compensated by the active circuit ($g=\beta$, $\beta_{\rm{e}}=0$), resulting in an extremely abrupt dip in $|S_{T}|$, as marked by an arrow in Fig.~\ref{fig2}(a). With the increase of the amplifier bias voltage, the gain coefficient exceeds the intrinsic damping rate, giving rise to an effective negative intrinsic damping rate of the cavity mode ($g>\beta$, $\beta_{\rm{e}}<0$). It is worth noting that the amplifier also affects the external damping rate of the cavity mode, as shown in Fig.~\ref{fig2}(c). However, its effect on the microwave transmission is not significant, and the transmitted signal appears as either amplification or attenuation, depending primarily on the compensation of the intrinsic dissipation (see Supplementary Material Sec.~II for details). By introducing the active circuit, the transmission amplitude becomes amplified around the cavity mode frequency, and the loss-compensated cavity is a crucial element for the subsequent realization of the nonreciprocal amplification.

With the amplification mechanism in hand, the next step is to construct the nonreciprocity of the system. Additionally, the YIG sphere is attached to the cavity. The magnon mode sustained by the YIG sphere is coherently and dissipatively coupled to the cavity photon mode. The nonreciprocity originates from the interference between the coherent and dissipative couplings, and the phase of the interference term is related to the direction of the microwave propagation, which leads to a direction-dependent complex coupling $\mathbb{C}_{R(L)}$. The magnon-photon coupling induces hybridized modes with complex frequencies
\begin{equation}\label{eq2}
\begin{split}
\tilde{\omega}_{\pm}=&\frac{1}{2}[\omega_{\rm{c}}+\omega_{\rm{m}}-i(\beta_{\rm{e}}+\alpha+\kappa_{\rm{e}}+\gamma)\pm\\
&\sqrt{\left[(\omega_{\rm{c}}-\omega_{\rm{m}})-i(\beta_{\rm{e}}-\alpha+\kappa_{\rm{e}}+\gamma)\right]^2+4\mathbb{C}^2}].
\end{split}
\end{equation}
Using the input-output theory {(see Supplementary Material Sec.~III for details)}, we obtain the transmission spectrum of the coupled system as
\begin{equation}\label{eq3}
	S_{\rm{21(12)}}=1+\frac{\kappa_{\rm{e}}}{i(\omega-\omega_{\rm{c}})-(\kappa_{\rm{e}}+\beta_{\rm{e}})+\frac{\mathbb{C}^2_{R(L)}}{i(\omega-\omega_{\rm{m}})-(\alpha+\gamma)}},
\end{equation}
where the complex coupling strength $\mathbb{C}_{R(L)}=M+iN$  between the cavity and magnon modes is direction-dependent as stated above, resulting in nonreciprocal transmission spectra, i.e., $S_{21}\neq S_{12}$. It is necessary to adjust the ratio between the coherent and dissipative coupling strengths for an optimal nonreciprocal transmission. Experimentally, we modify the complex coupling strength $\mathbb{C}_{R(L)}$ by fixing the height of the YIG sphere relative to the planar device at a distance of $D=1$ mm and then precisely controlling its position in the x-y plane.

To clearly show the existing region of the nonreciprocal transmission exists and reveal its origin in our device, here we place the YIG sphere at three different positions relative to the cavity, as shown in Fig.~\ref{fig3}(a). We depict the corresponding $|S_{21}|$ and $|S_{12}|$ mappings versus the field detuning $\Delta_{\rm{m}}=\omega-\omega_{\rm{m}}$ and the probe frequency detuning $\Delta_{\rm{c}}=\omega-\omega_{\rm{c}}$ in Figs.~\ref{fig3}(c)-(e), respectively. When the YIG sphere is at position 1, the two hybridized modes shown in Fig.~\ref{fig3}(c) exhibit level repulsion, a characteristic of coherent magnon-photon coupling~\cite{Hans-13,Zhang-14,Tabuchi-14}. Using Eq.~(\ref{eq2}) to fit the experimental results, we find that at position 1 the real part of the complex coupling strength $M=|\rm{Re}(\mathbb{C})|$ is significantly larger than the imaginary part $N=|\rm{Im}(\mathbb{C})|\approx 0$, as depicted by the blue dot in Fig.~\ref{fig3}(b). Nonreciprocity cannot be observed in this instance because the interference term between the coherent and dissipative couplings is negligibly small. As observed in the left ($|S_{21}|$) and right panels ($|S_{12}|$) of Figure~\ref{fig3}(c), transmission mappings are nearly identical.

When the YIG sphere is placed at position 2, the mode hybridization as shown in Fig.~\ref{fig3}(d) is quite distinct from level repulsion but similar to level attraction, indicating the coexistence of dissipative and coherent magnon-photon couplings~\cite{WangY-19}. As indicated by the orange dot in Fig.~\ref{fig3}(b), when the YIG sphere is placed at position 2, the coherent and dissipative coupling strengths are comparable. The interference between coherent and dissipative couplings then plays a crucial role.  Consequently, the nonreciprocity is evident in Fig.~\ref{fig3}(d), where $|S_{21}|\neq|S_{12}|$ is observed at various detunings. As an illustration, we plot the nonreciprocal transmission at the arrow-marked field detuning in Fig.~\ref{fig3}(d). As depicted in the inset of Fig.~\ref{fig3}(b), red and blue curves correspond to $|S_{21}|$ and $|S_{12}|$, respectively.

The coupling strength $\mathbb{C}$ tends to be zero [gray dot in Fig.~\ref{fig3}(b)] when the YIG sphere is placed at position 3, which is far from the cavity and transmission line. Conceivably, nonreciprocal transmission does not occur in this case either. The measured transmission mappings of $|S_{21}|$ and $|S_{12}|$ are shown in Fig.~\ref{fig3}(e). Through the measurements at the positions above, we can find that the coexistence of coherent and dissipative couplings is very crucial for the emergence of nonreciprocal transmission. In addition, to obtain the maximum nonreciprocal response, the coherent and dissipative coupling strengths should be comparable.

Combining the transmission amplification caused by the dissipation-compensated cavity and the nonreciprocity of the coupled system, we achieve the unidirectional amplification and isolation, as shown in Fig.~\ref{fig4}. In order to quantify isolation, the difference between the rightward ($S_{21}$) and leftward ($S_{12}$) transmission amplitudes is extracted and plotted as $20{\log}_{10}|S_{21}/S_{12}|$. Here we use its absolute value as the isolation ratio (Iso.) of the system. The experimentally observed isolation ratios at different amplifier bias voltages ($V$=0, 4.8, 5 V) are plotted versus $\Delta_{\rm{m}}$ and $\Delta_{\rm{c}}$ in Figs.~\ref{fig4}(a)-(c). As the bias voltage of the amplifier rises, the intrinsic damping rate of the hybrid system transits from the positive to negative, which means the generation of a net gain in the system (see Supplementary Material Sec.~IV). The maximum isolation ratio increases as more signal gain is supplied to the system. On the premise of a high level of isolation, it is crucial to find out the optimal operating point in order to achieve a relatively large unidirectional signal amplification. We choose the field detuning position marked by the gray planes in Figs.~\ref{fig4}(a)-(c) as examples, and plot the $|S_{21}|$ and $|S_{12}|$ spectra in Figs.~\ref{fig4}(d)-(f), respectively. The circle points are the experimental results, while the solid curves are the theoretical results obtained using Eq.~(\ref{eq3}), which are in good agreement. The microwave transmission depicted in Fig.~\ref{fig4}(d) is nonreciprocal without amplification, because the bias voltage of the amplifier is set to zero. In Figs.~\ref{fig4}(e) and \ref{fig4}(f), the unidirectional amplification is realized when bias voltage is applied to the amplifier. At the working frequency of $\omega/2\pi=3.003$~GHz, the rightward propagating microwave can be amplified by 11.5~dB, whereas the leftward propagating microwave can be attenuated by 34.7~dB, giving an isolation ratio of 46.2~dB. Fitting these spectra with Eq.~(\ref{eq3}), we obtain the propagation direction-dependent complex coupling strengths of the system, which are $\mathbb{C}_{R}=(5.66-4.3i)$ MHz and $\mathbb{C}_{L}=(5.6-2.74i)$ MHz.

%\section{Conclusions}
In conclusion, we have designed a novel nonreciprocal device that performs unidirectional amplification and reserve isolation simultaneously. The device consists of an active circuit with an embedded amplifier, a passive cavity made by SRR, and a YIG sphere. The unidirectional amplification of the microwave signal is accomplished by combining the nonreciprocity of the cavity magnonic system with the conventional active feedback of the gain medium. By adjusting the bias voltage of the amplifier in the active feedback circuit, we can compensate for loss of the passive cavity mode and convert it to a gain regime. Meanwhile, the coupling strength between the cavity photon mode and the magnon mode is controlled by the relative position between the device and the YIG sphere. By balancing the coherent and dissipative couplings, the interference between them can result in nonreciprocity in our system. By optimization via amplifier bias voltage and photon-magnon coupling strength, we are able to achieve microwave transmission amplification of 11.5 dB in one propagation direction and attenuation of 34.7 dB in the opposite propagation direction. Such a unidirectional amplification device may have promising applications in quantum information technologies, as it can both amplify the weak signal leaking from the principal quantum system and isolate the back-scattered noise from the readout electronics. This design of integrating two functions in just one device will provide advantages for the construction of large-scale information networks.

\begin{acknowledgments}
We thank Zi-Qi Wang for helpful discussion. This work is supported by the National Key
Research and Development Program of China (Grant No.~2022YFA1405200), the National Natural Science Foundation of China (Grants No.~92265202, No.~11934010, and No.~12174329), the Fundamental Research Funds for the Central Universities (Grant No.~2021FZZX001-02).
\end{acknowledgments}

\section*{Data Availability Statement}
The data that support the findings of this study are available from the corresponding author upon reasonable request.

%\begin{thebibliography}{999}
\providecommand{\noopsort}[1]{}\providecommand{\singleletter}[1]{#1}%
%

%\end{thebibliography}

\end{document}